\documentclass[12pt]{article}

\begin{document}

\author{James T. Wheeler}
\title{Extended Conformal Symmetry}
\maketitle

\begin{abstract}
We show that the grading of fields by conformal weight, when built into the
initial group symmetry, provides a discrete, non-central conformal extension
of any group containing dilatations. We find a faithful vector
representation of the extended conformal group and show that it has a
scale-invariant scalar product and satisfies a closed commutator algebra.
The commutator algebra contains the infinite Heisenberg and Virasoro
algebras. In contrast to the usual treatments of scale invariance, covariant derivatives and gauge transformations automatically incorporate the
correct conformal weights when the extended symmetry is gauged.
\end{abstract}

\section{Introduction}

Scale invariance provides an successful class of physical (string) models
in 2-dim. One element of the success in 2-dim is the infinite character of
the conformal Lie algebra in that dimension. However, as we show here, the
standard formulation of scale invariance introduced by Weyl \cite{Weyl} is
incomplete. By studying scale \textit{covariance} - the grading of fields by
conformal weight implicit in Weyl or conformal geometries - we find a
noncentral, conformal extension of any Lie group containing dilatations. We
find that the infinite Heisenberg and Virasoro algebras are natural adjuncts
of scale covariance in any dimension. When the extended groups are gauged
the resulting spaces possess a natural commutator product and a
scale-invariant inner product.

Initially motivated by the assignment of weights to classical fields, the
use of scale covariance is transformed to a necessity by the well-known
result that in quantum field theory, scale-invariant theories spontaneously
develop scales. Thus, we build a conformally \textit{covariant} theory based on conformal
or other scaling symmetries. We start with the observation that it is not
only the \textit{units} of length that are arbitrary. In addition, at the
outset we can assign an arbitrary conformal \textit{weight} to the
coordinates -- then the scale invariance of the action fixes the weights of
other fields. When this freedom is described as a symmetry we are led to
introduce a set of operators that change the assigned weight. \textit{The
full collection of such operators forms a noncentral, discrete extension of
the conformal group which satisfies any of the standard definitions of
conformal transformations} (i.e., preserve angles, preserve ratios of
infinitesimal displacements, preserve light cones \cite{Barut}, \cite
{Hughston and Hurd}).

Extended scaling groups and their gauge theories have a range of novel
properties. First, they fill some gaps in the previous formulation. In
the past, the covariance of fields has been managed by inserting the
conformal weight by hand into gauge transformations and the covariant
derivative. But in extended conformal gauge theories, the geometric
structure automatically recognizes the weights of fields. This provides a
check on our development, indicating the correctness of the extension.

What is more interesting than completing Weyl's picture, however, is the
potential usefulness of the new structures for building new field theory
models. Extended scaling groups are Lie groups with infinitely many
connected components. Tangent vectors to the group manifold are both
homothetic tensors and elements of an infinite dimensional vector space,
which obey a commutator algebra containing the infinite Heisenberg algebra,
the Virasoro algebra, homothetic algebras and Kac-Moody algebras as
sub-algebras. These vectors also have an invariant scalar product which is
positive definite on the conformally self-dual subspace.

We make use of a recent advance in conformal gauge theory in $n$-dim to
display some of these properties of the extended symmetry. In the past,
conformal gaugings (\cite{Romao+Ferber+Freund}-\cite{Townsend+van
Nieuwenhuizen}) treated the special conformal transformations as additional
symmetries of the final gauged geometry. But \textit{biconformal gauging}
requires the special conformal transformations together with the
translations to span the base manifold, giving a $2n$-dim space with $n$-dim
Weyl symmetry group \cite{New Conformal Gauging Paper}. Because the
biconformal gauging has a dimensionless volume form, it is possible to write
a scale-invariant action linear in the biconformal curvatures. This action
leads to two involutions of the $2n$-dim space which show it to be spanned
by a Ricci-flat, $n$-dim Riemannian spacetime together with a flat $n$-dim
Riemannian spacetime \cite{WWI}. Thus, higher-dim conformal symmetry is
consistent with vacuum general relativity in a straightforward way, without
requiring a quadratic action or compensating fields. Recently, these results
have been extended to include certain matter sources (\cite{WWII},\cite
{WWIII}). Below we illustrate the use of extended conformal symmetry by
performing a biconformal gauging of the extended conformal group.

\smallskip The layout of the paper is as follows. In the next section, we
derive the form of extended scaling groups from the usual representation for
dimensionful fields. Then, in Sec.(3), we develop a vector representation on
which the extended group acts effectively. The vector representation is
isomorphic to the space of tangent vectors to the extended group manifold.
We show that these vectors satisfy a commutator algebra containing the
infinite Heisenberg and Virasoro algebras and define a class of inner
products. Finally, in Sec.(4), we indicate a few of the consequences of the
structures of Sec.(3) for field theories based on the extended conformal
group. Sec.(5) contains a brief summary.

\section{Extended Scale Invariance}

In this section and the next, we establish our main results. Here we define
the extended conformal group, find its center and find their quotient group.
The quotient group acts transitively and effectively. In the next section,
we find a vector representation of the group and show that the vectors close
under commutation and have an invariant, indefinite scalar product.

To build the group, we first note that nothing physical is changed if we
assign an arbitrary conformal weight to the spacetime coordinates, adjusting
the weights of other fields accordingly. For simplicity, we take the
possible weight assignments to lie in the set $Z$\ of integers. This much is
necessary to account for our ability to step the conformal weight up or down
arbitrarily many times by integrating and differentiating fields.

Now we start with the standard definition: \textit{The extended conformal
group, $\mathcal{C}_{E},$} \textit{is the group of transformations that
preserves angles.} We understand this definition to include transformations
of conformal weight, so, in addition to the usual conformal transformations, 
$\mathcal{C}_{E}$ includes an infinite discrete part which formalizes the
conformal grading. The development is as follows.

Suppose $\mathcal{G}$ is a Lie symmetry group including the $1$-dim
dilatational subgroup, $e^{\lambda D},$ with generator $D.$ Then the
generators $\{G_{A}\}=\{D,G_{\alpha }\}$ of the Lie algebra, $\mathcal{L}%
_{g},$ of $\mathcal{G}$ satisfy commutation relations of the form 
\begin{eqnarray}
\lbrack D,G_{\alpha }] &=&c_{0\alpha }^{\quad \beta }G_{\beta }+c_{0\alpha
}^{\quad 0}D \\
\lbrack G_{\alpha },G_{\beta }] &=&c_{\alpha \beta }^{\quad \gamma
}G_{\gamma }+c_{\alpha \beta }^{\quad 0}D
\end{eqnarray}
Examples for $\mathcal{G}$ include the homogeneous and inhomogeneous
homothetic (or Weyl) groups and the conformal group. We wish to study tensor
field representations $\phi \in \Phi $ of $\mathcal{G}$ which in addition to
their transformation properties under $\mathcal{G}$%
\begin{equation}
\phi ^{\prime }(x^{\prime })=e^{\lambda ^{A}G_{A}}\phi (x)
\end{equation}
are assigned geometric units from the set $L=\{(length)^{m}|m\in Z\}.$ This
graded field representation therefore takes the product form 
\begin{equation}
\phi (x)(length^{k})=\phi (x)\ l^{k}\in \Phi \otimes L
\end{equation}
We seek an extension, $\mathcal{G}_{E},$ of the original group $\mathcal{G},$
that acts on the representation $\Phi \otimes L.$ The extension should
include a subgroup, $\mathcal{J}$, of operators $J_{\Sigma }:L\rightarrow L.$
Thus, $\mathcal{J}$ is a subgroup of the automorphism group of $L;$ in
addition we ask for $\mathcal{J}$ to be closed under the action of $\mathcal{%
G}$. Since we expect, for example, Lorentz transformations or translations
to commute with changes of assigned conformal weight, this amounts to
closure under dilatations, $e^{\lambda D}\mathcal{J}e^{-\lambda D}\subseteq 
\mathcal{J}$. The largest such group is easily seen (see Appendix A) to be
the set $\{J_{k}|J_{k}l^{m}=l^{m+k}\}$ satisfying $e^{\lambda
D}J_{k}e^{-\lambda D}=e^{-\lambda k}J_{k}.$ It follows that 
\begin{eqnarray}
\lbrack J_{k},J_{m}] &=&0 \\
\lbrack J_{m},D] &=&mJ_{m}  \label{J weight} \\
\lbrack J_{m},G_{\alpha }] &=&0 \\
J_{k}J_{m} &=&J_{k+m}  \label{J product}
\end{eqnarray}

We form a group containing both $D$ and $\{J_{k}\}$ by writing 
\begin{equation}
h(\alpha ,k,\lambda )=e^{\alpha }J_{k}e^{\lambda D}
\end{equation}
Then the group product is 
\begin{eqnarray}
h(\alpha ,k,\lambda )h(\beta ,m,\gamma ) &=&e^{\alpha +\beta -\lambda
m}J_{k+m}e^{(\lambda +\gamma )D}  \label{Mult law} \\
&=&h(\alpha +\beta -\lambda m,k+m,\lambda +\gamma )
\end{eqnarray}
The identity element is $h(0,0,0)=J_{0},$ and the inverse to $h(\alpha
,k,\lambda )$ is $h(-\alpha +\lambda k,-k,-\lambda ).$ Notice that the
positive real factor $e^{\alpha }$ is necessary in order to accomodate the
factor $e^{-\lambda m}$ that arises in commuting $e^{\lambda D}$ and $J_{m}$
into standard form after taking a product.

Next, we extend the dilatation factor, $e^{\lambda D},$ to include the rest
of the group $\mathcal{G}$. As noted above, the remaining generators $%
G_{\alpha }$ may be expected to commute with $J_{k}$. Therefore a general
element $g\in \mathcal{G}_{E}$ takes the form 
\begin{equation}
g(\alpha ,k,\lambda ,\lambda ^{\alpha })=e^{\alpha }J_{k}\ e^{\lambda
D+\lambda ^{\alpha }G_{\alpha }}  \label{Extended group element}
\end{equation}
where $\lambda D+\lambda ^{\alpha }G_{\alpha }$ is a general element of $%
\mathcal{L}_{g}$. We easily check that $\mathcal{G}_{E}$ is a Lie group. In
particular, the inverse of $g$ is given by 
\begin{equation}
g^{-1}=e^{-\alpha }e^{-\lambda D-\lambda ^{\alpha }G_{\alpha
}}J_{-k}=e^{-\alpha -\lambda k}J_{-k}\ e^{-\lambda D-\lambda ^{\alpha
}G_{\alpha }}\in \mathcal{G}_{E}
\end{equation}

The Lie algebra, $\mathcal{L}_{g_{E}},$ of $\mathcal{G}_{E}$ is $\mathcal{L}%
_{g}$ extended by the identity to include the necessary factor $e^{\alpha }:%
\mathcal{L}_{g_{E}}=\mathcal{L}_{g}\oplus \mathbf{1}.$ The $J_{k}$ operators
which change conformal weights give the group an infinite number of distinct
connected components. Clearly, each of these components is a manifold of
dimension ($\dim \mathcal{G}+1)$ which is in $1-1$ correspondence with $%
\mathcal{G\otimes }R^{+}$. Thus, as a manifold, $\mathcal{G}_{E}$ is
homeomorphic to the direct product, $\mathcal{G\otimes }R^{+}\mathcal{%
\otimes }\mathbf{Z.}$

In order to construct a gauge field theory based on $\mathcal{G}_{E},$ we
study the adjoint action of $\mathcal{G}_{E}$ on the manifold $\mathcal{M}=%
\mathcal{G}_{E},$ seeking the maximal effective subgroup. This will be $%
\mathcal{G}_{E}$ modulo its center, where the center of $\mathcal{G}_{E}$ is
the set 
\begin{equation}
K=\{g|gpg^{-1}=p,\forall p\in \mathcal{G}_{E}\}=\{e^{\alpha }\}
\end{equation}
Notice that the elements $J_{k}$ do \textit{not} lie in the center because
for a general $p=J_{m}\ e^{\alpha +\lambda D+\lambda ^{\alpha }G_{\alpha
}}\in \mathcal{M}$ we have 
\begin{equation}
J_{k}pJ_{-k}=J_{k}(J_{m}\ e^{\alpha +\lambda D+\lambda ^{\alpha }G_{\alpha
}})J_{-k}=e^{\lambda k}p\neq p
\end{equation}
However, the projective subgroup $\mathcal{G}_{E}/K$ is isomorphic to the
direct product $\mathcal{G}\otimes \mathcal{J}$ because $J_{k}$ and $g_{c}=$ 
$e^{\lambda D+\lambda ^{\alpha }G_{\alpha }}$ now commute in $\mathcal{G}%
_{E}/K$: 
\begin{equation}
J_{k}g_{c}=J_{k}e^{\lambda D+\lambda ^{\alpha }G_{\alpha }}=e^{k\lambda
}g_{c}J_{k}\cong g_{c}J_{k}
\end{equation}
The quotient $\mathcal{G}_{E}/K$ is the maximal effective subgroup. The
quotient introduces central charges into both the Lie algebra of $\mathcal{G}
_{E}/K,$ and into the commutator algebra of the extended representation and
tangent space. The central charges for $\mathcal{G}_{E}/K$ depend only on
the original Lie algebra of $\mathcal{G}.$ Those of the tangent space are
discussed below.

We note also that $\mathcal{G}_{E}/K$ is transitive, since any two elements $%
p=J_{k}g_{c1}$ and $q=J_{m}g_{c2}$ are connected by the element $%
r=J_{m-k}g_{c2}g_{c1}^{-1}:$%
\begin{equation}
rp=J_{m-k}g_{c2}g_{c1}^{-1}J_{k}g_{c1}\cong J_{m}g_{c2}=q
\end{equation}

We now find a representation for $\mathcal{G}_{E},$ which turns out to be
isomorphic to the tangent space, $T\mathcal{G}_{E}$ of $\mathcal{G}_{E}.$

\section{The extended representation space}

In this section, we continue to study properties of $\mathcal{G}_{E},$
finding a faithful vector representation for $\mathcal{G}_{E}$ and
developing its properties. Normally, the Lie algebra of a Lie group provides
an adequate vector representation for a Lie group. However, when we wish to
represent discrete symmetries this representation is not faithful and we
must look at infinitesimal transformations in \textit{each }connected
component, rather than just a neighborhood of the identity. To illustrate
this point, we first consider the trivial example of $O(3),$ including
parity. Then we apply the technique to $\mathcal{G}_{E}.$

For $O(3),$ a general group element may be written in the form $g=P_{\alpha
}e^{\frac{1}{2}\lambda ^{i}M_{i}}$ where $\alpha \in \{+,-\},$ $P_{+}=%
\mathbf{1},P_{-}=-\mathbf{1}$ and the $M_{i}$ generate rotations. The $%
P_{\alpha }$ have product $P_{\alpha }P_{\beta }=P_{\alpha \times \beta }.$
An element of the Lie algebra is simply $v=v^{i}M_{i},$ which is
insufficient to represent the action of parity. Instead, we expand about
each connected component to find six generators, $P_{+}M_{i}$ and $%
P_{-}M_{i},$ and a representation of the form 
\begin{equation}
v=v_{+}^{i}(P_{+}M_{i})+v_{-}^{i}(P_{-}M_{i})
\end{equation}
The enlarged vector space is sufficient to span such common indefinite
parity combinations as 
\begin{equation}
\vec{y}=\vec{w}+\vec{u}\times \vec{v}=(\vec{u}\times \vec{v}%
)^{k}(P_{+}M_{k})+w^{k}(P_{-}M_{k}).
\end{equation}
Notice that the basis vectors form a commutator algebra under the usual
Leibnitz rule for the commutator of a product. For example, the commutator
of two vectors in the $P_{-}$ sector appropriately lie in the $P_{+}$ sector
since 
\begin{equation}
\lbrack v,w]=v^{i}w^{j}[P_{-}M_{i},P_{-}M_{j}]=(\vec{v}\times \vec{w}%
)^{k}P_{+}M_{k}
\end{equation}

Similarly, to find a faithful representation of $\mathcal{G}_{E}$ we
consider the expansion of $\mathcal{G}_{E}$ about each $J_{k}.$ We find 
\begin{equation}
e^{\alpha }J_{k}\ e^{\lambda D+\lambda ^{\alpha }G_{\alpha }}\ \approx\
J_{k}(1+\alpha 1+\lambda D+\lambda ^{\alpha }G_{\alpha })
\end{equation}
so that the representation has the basis 
\begin{equation}
A=(J_{k}\mathbf{1},J_{k}D,J_{k}G_{\alpha })\equiv (J_{k},L_{k},G_{\alpha
}^{k})
\end{equation}
where we have defined 
\begin{eqnarray}
L_{k} &\equiv &J_{k}D \\
G_{\alpha }^{k} &\equiv &J_{k}G_{\alpha }
\end{eqnarray}
It is straightforward to show that vectors of the form 
\begin{equation}
v=\sum_{k=-\infty }^{\infty }\left( v^{k}J_{k}+v^{0k}L_{k}+v^{\alpha
k}G_{\alpha }^{k}\right) \in V
\end{equation}
give a faithful representation of $\mathcal{G}_{E}.$ In fact, this
representation is isomorphic to the tangent space to $\mathcal{G}_{E}.$ The
components of $v$ have conformal weights $(k,k,k+w_{\alpha })$ where $%
w_{\alpha }$ is the weight of $G_{\alpha }.$

Since $[J_{m},J_{n}]=0$ and because of the form of the product $J_{k}J_{m}$
given in eq.(\ref{J product}), this space also admits a commutator product
between vectors, again defined using $[AB,C]=A[B,C]+[A,C]B$. This algebra is
generally \textit{not} a Lie algebra because the Jacobi identities fail.
Instead, we have Jacobi relations, given in Appendix B. All of the
non-Jacobi terms are proportional to $c_{\alpha \beta }^{\quad 0}$ or $%
c_{0\alpha }^{\quad 0},$ so for the case of homogeneous or inhomogeneous
homothetic algebras, the usual Jacobi identities do hold. Recalling that the
projective quotient also introduces central charges into $\mathit{V}$, the
algebra of $\mathit{V}$ becomes

\begin{eqnarray}
\lbrack G_{\alpha }^{k},G_{\beta }^{m}] &=&c_{\alpha \beta }^{\quad \gamma
}G_{\gamma }^{k+m}+c_{\alpha \beta }^{\quad 0}L_{k+m}+c_{\alpha \beta
}\delta _{k+m}^{0}  \label{G commutator} \\
\lbrack L_{k},G_{\alpha }^{m}] &=&(c_{0\alpha }^{\quad \beta }-m\delta
_{\alpha }^{\beta })G_{\beta }^{k+m}+c_{0\alpha }^{\quad 0}L_{k+m} \\
\lbrack L_{k,}L_{m}] &=&(k-m)L_{k+m}+ak(k^{2}-1)\delta _{k+m}^{0} \\
\lbrack J_{m},L_{k}] &=&mJ_{k+m}+b_{k}\delta _{k+m}^{0} \\
\lbrack J_{k},J_{m}] &=&ck\delta _{k+m}^{0}
\end{eqnarray}
The algebra has several readily identifiable properties.

\begin{enumerate}
\item  The subalgebra generated by $\{G_{\alpha }^{0},L_{0},J_{0}=\mathbf{1}%
\}$ is the original extended Lie algebra of $\mathcal{G}_{E},$ while $%
\{G_{\alpha }^{0},L_{0}\}$ is the Lie algebra of $\mathcal{G}$.

\item  The subalgebra generated by $\{J_{k}\}$ is the infinite Heisenberg
algebra.

\item  The subalgebra generated by $\{L_{k}\}$ is the Virasoro algebra.

\item  For any proper Lie subalgebra of $\mathcal{L}_{g},$ with basis $%
\{H_{\beta }\}\subset \{G_{\alpha }\},$ the set $\{H_{\beta }^{k}\}$ is a
basis for the associated Kac-Moody algebra. For example, when $\mathcal{G}$
is the conformal group, eq.(\ref{G commutator}) includes the
Poincar\'{e}-Kac-Moody algebra. Notice that the commutator algebra does 
\textit{not} contain the Kac-Moody algebra of $\mathcal{G}$ because of the
nontrivial commutator for the $L_{k}.$ This commutator is nontrivial
precisely because the extension is noncentral, i.e., the dilatation
generator $D$ measures the weight of $J_{k}$ the same way it measures the
weight of $G_{A}.$
\end{enumerate}

\smallskip

We may also define a class of indefinite, weight-$m$ scalar products for $%
\mathit{V}$ whenever the original group has a nontrivial Killing metric, $%
K_{AB}=c_{EA}^{\quad F}c_{FB}^{\quad E}.$ Using the adjoint representation
the Lie algebra of $\mathcal{G}$ we have 
\begin{equation}
K_{AB}=tr(G_{A}G_{B})
\end{equation}
so that if we define 
\begin{equation}
Tr(J_{k}G_{A})=\delta _{k}^{0}\ tr(G_{A})
\end{equation}
we have 
\begin{eqnarray}
\langle v,w\rangle _{m} &\equiv &Tr(vJ_{-m}w) \\
&=&\sum_{k}v^{A,k}w^{B,m-k}\ K_{AB}=\langle w,v\rangle _{m}
\end{eqnarray}
where we have used $tr(G_{A})=0.$ The result has weight $m$ because we
require the weight of $v=v^{m}J_{m}$ to be zero. Of greatest interest is the
scale-invariant case, $m=0,$%
\begin{equation}
\langle v,w\rangle \equiv \langle v,w\rangle
_{0}=Tr(vw)=\sum_{k}v^{Ak}w^{B,-k}K_{AB}
\end{equation}

We can find a subspace on which he $0$-weight inner product defines a norm.
Let the conformal dual of a vector be defined as 
\begin{equation}
\bar{v}\equiv \sum v^{kA}J_{-k}G_{A}
\end{equation}
and define $v$ to be self-dual if there exists a gauge in which $v=\bar{v}.$
Then a gauge-invariant norm is given on the space of self-dual vectors by 
\begin{equation}
\parallel v\parallel ^{\ 2}\equiv \langle v,v\rangle
=\sum_{k}v^{kA}v^{kB}K_{AB}
\end{equation}
In general, $\parallel v\parallel ^{\ 2}$ shares the signature of $K_{AB}.$
For self-dual vectors of the form $v_{v}=v^{k}J_{k}$ (vertical on the bundle
defined below) the norm is positive definite.

As an example of the use of the extended conformal group, we now consider
its biconformal gauging.

\section{Gauging the extended conformal group}

The connected component of the extended conformal group only differs from
the conformal group through the presence of the positive real factor, $%
e^{\alpha },$ and this part is factored out to produce an effective group
action. Therefore, there is no difference between the \textit{local}
structure of extended biconformal gauge theory and the biconformal gauging
described in (\cite{New Conformal Gauging Paper}-\cite{WWIII}). For this reason, we give only a brief summary of the biconformal space, then move directly to some properties of graded tensor fields.

We first consider the action of $\mathcal{G}_{E}/K$ on $\mathcal{M}=\mathcal{%
G}_{E}/K.$ As just noted, the local structure of $\mathcal{M}$ is described
by the original Lie algebra of $\mathcal{G}.$ This means that even though $%
\mathcal{M}$ has multiple connected components, the connection is still a $%
\mathcal{G}$-valued $1$-form on $\mathcal{M}.$ Each of the connected
components therefore shares the same curvature, and $\mathcal{M}$ is
homeomorphic to a direct product, $J\otimes \mathcal{M}_{0}.$ The $\mathcal{G%
}$-valued connection is sufficient to provide a unique $\mathcal{G}$ mapping
along any given curve between any two points on the same connected
component, while the discrete operators $J_{k}$ map uniquely from component
to component. Thus, the direct product allows us to define a $\mathcal{G}%
_{E}/K$ action on $\mathcal{M}.$ Moreover, the tangent bundle to $\mathcal{M}
$ includes a copy of the tangent space associated with each connected
component, giving an infinite-dimensional irreducible vector representation
of $\mathcal{G}_{E}/K$. The direct product structure of the base manifold, $%
J\otimes \mathcal{M}_{0},$ means the manifold itself may be treated as a
trivial bundle with projection $\pi :J_{k}\rightarrow \mathbf{1.}$ The
tangent space $T\mathcal{G}_{E}$ may be divided into horizontal and vertical
vector spaces using this projection.

Our notation follows that of refs \cite{New Conformal Gauging Paper} and 
\cite{WWI}, and is based on $O(n,2)$. The fibre bundle is given by the
quotient $\mathcal{C}/\mathcal{W}$, with the connection $1$-form 
\begin{equation}
\mathbf{\omega =}\frac{1}{2}\mathbf{\omega }_{b}^{a}M_{a}^{b}+\mathbf{\omega 
}_{0}^{0}D
\end{equation}
where the $M_{b}^{a}$ generate Lorentz transformations. The biconformal base
manifold is spanned by the $2n$ $1$-forms $(\mathbf{\omega }_{0}^{a},\mathbf{%
\omega }_{a}^{0})$. Because these basis forms have opposite scaling weights,
biconformal geometry has a scale invariant volume form, and allows us to
write a scale invariant action linear in the curvature tensors without the
use of compensating fields. The resulting field equations, subject to a
constraint of minimal torsion, lead to a foliation of the $2n$-dim space by $
n$-dim Riemannian spacetimes satisfying the vacuum Einstein equation. The
theory therefore makes close contact with general relativity. The structure
equations, curvatures and gravitational field equations are reported in
detail in \cite{WWI}.

We now consider tensor fields on such a biconformal geometry. The effect of
such matter fields on the results of \cite{WWI} are under separate
investigation \cite{WWIII}, so we will limit our focus to a few basic
properties of graded matter fields in flat space. For tangent vectors we
have 
\begin{equation}
v=v^{m}J_{m}+v_{m}^{a}P_{a}^{m}+v_{a}^{m}K_{m}^{a}
\end{equation}
where the generators $P_{a}$ and $K^{a}$ are for translations and
co-translations, respectively. The covariant derivative is given by 
\begin{eqnarray}
\mathbf{D}v &=&\mathbf{d}v+[\mathbf{\omega },v] \\
&=&(\mathbf{d}v^{m}+m\mathbf{\omega }_{0}^{0}v^{m})J_{m} \\
&&+(\mathbf{d}v_{m}^{a}-v_{m}^{c}\mathbf{\omega }_{c}^{a}+(m+1)\mathbf{%
\omega }_{0}^{0}v_{m}^{a})P_{a}^{m} \\
&&+(\mathbf{d}v_{a}^{m}+\mathbf{\omega }_{d}^{c}v_{c}^{m}+(m-1)\mathbf{%
\omega }_{0}^{0}v_{a}^{m})K_{m}^{a}
\end{eqnarray}
which shows that the scale-covariant derivative, with appropriate weights,
emerges correctly. This result confirms our claim at the start, that the
extended conformal group completes Weyl's description of scale invariance.
We can now differentiate arbitrary weight fields correctly without inserting
the weights by hand.

In addition to providing a consistent formalism, the extended group has interesting field theoretic properties. Consider the dynamics of a general weight vector, $v=v^{m}J_{m}.$ We easily write a massive, scale-invariant action for $v,$ using the invariant inner product: 
\begin{equation}
\frac{1}{2}\sum_{n=-\infty }^{\infty }\int (\langle \mathbf{D}v,\mathbf{D}%
v\rangle +\ \langle \mathrm{m}^{2}v,v\rangle )\mathbf{\Phi }
\end{equation}
where $\langle \mathbf{D}v,\mathbf{D}v\rangle =\sum
K^{AB}D_{A}v^{n}D_{B}v^{-n}$ and $\mathbf{\Phi }$ is the dimensionless
biconformal volume element. Then weight of the mass is taken as $-1,$ so
that $\langle \mathrm{m}^{2}v,v\rangle =\sum \mathrm{m}^{2}v^{n+2}v^{-n}.$
Neglecting gravitational effects, the field equations are 
\begin{equation}
K^{AB}D_{A}D_{B}v^{n}=\ \mathrm{m}^{2}v^{n+2}  \label{Vert field eq.}
\end{equation}
where indices $A,B,\ldots $ run over the full basis, $\omega ^{A}=(\omega
_{0}^{a},\omega _{a}^{0})$ and $K^{AB}$ is the projection of the conformal
Killing metric to the base space, $K^{ab}=K_{ab}=0,K_{\quad b}^{a}=K_{b}^{\quad a}=\delta_{b}^{a}$. Substituting the form of $K^{AB}$ in the
expressions above gives 
\begin{equation}
K^{AB}D_{A}D_{B}v^{n}=(D^{a}D_{a}+D_{a}D^{a})v^{m}
\end{equation}

While eq.(\ref{Vert field eq.}) appears to be a straightforward classical
wave equation, there are two important differences. First, we note that the
presence of the mass term couples component fields of different conformal
weight. This means that if the theory is not to break conformal invariance,
it must be massless. If we keep the mass we necessarily produce mixing
between different weight fields. Of course, such mixing is also produced by
generic potentials, $U(v).$

To see the second difference, recall the commutator algebra satisfied by
vertical tangent vectors. For simplicity, let the spacetime be flat and let $%
\pi _{m}$ be the canonically conjugate momentum to $v^{m}.$ Then 
\begin{equation}
\pi _{m}=\frac{\partial \mathcal{L}}{\partial \left( \frac{\partial v^{m}}{%
\partial x^{0}}\right) }=\frac{\partial v^{-m}}{\partial y_{0}}=\partial
^{0}v^{-m}
\end{equation}
Notice that, for scale-invariant actions, canonical conjugacy and conformal
conjugacy always coincide. For $v^{m}$ and $\pi _{m},$ the commutator
algebra with central charges gives 
\begin{equation}
\lbrack v,\pi ]=amv^{m}\partial ^{0}v^{-m}  \label{Commutator}
\end{equation}
Therefore, we have a nonvanishing commutation relation between a field and
its conjugate momentum. It is shown in \cite{WWII} that the term on the
right is proportional to one component of the Weyl vector. The commutator
arises because of the central charge, $a.$ Recall that the central charge, $
a,$ is a necessary consequence of the exponential factor in the original
group and the demand that the group act effectively.

The commutator in eq.(\ref{Commutator}) is quadratic in the fields, and
therefore of the same order as the source terms for the gravitational sector
of the geometry. For this reason, we cannot explore the consequences of eq.(%
\ref{Commutator}) further here -- the only solution available is the $m=0$
case of a scalar field (see \cite{WWIII} for a description of this solution,
and \cite{WWII} for the mathematical details). More general classes of
solution are under active investigation.

What can be said at this point is that in order to maintain the commutator
algebra of the tangent space while using the vectors for field theory, we
must use something akin to the techniques of quantum field theory. Indeed, $%
v $ is already in the form of the usual Hiesenberg operator expansion for a
quantum field$.$ Thus, in addition to successfully formalizing certain
details of scale invariance, the structures arising from the extended
conformal group have unexpected properties which might shed some insight
onto our understanding of quantum systems.

Before concluding, we note one further possible form of the scalar field
action, in which we take the norm instead of the inner product: 
\begin{equation}
\frac{1}{2}\sum_{n=-\infty }^{\infty }\int \Vert \mathbf{D}v\Vert ^{2}%
\mathbf{\Phi }
\end{equation}
Here we define $\Vert \mathbf{D}v\Vert ^{2}\equiv K^{AB}\langle D_{A}\bar{v}
,D_{B}v\rangle .$ In this case, we can add a mass term $\mathrm{m}^{2}\Vert
v\Vert ^{2}\mathbf{\Phi }$ only for fields $v^{k}$ of definite weight $k=1,$
but for this one case there is no mixing of conformal weights.

\section{Conclusion}

We have shown that our freedom to assign an arbitrary conformal weight to
spacetime coordinates leads to a noncentral, discrete extension of the
conformal group. This discrete extension applies to any Lie group
containing dilatations. We examined the properties of the resulting extended
scaling groups and their gauge theories.

We showed the presence of central charges in the extended algebra. \textit{A
priori}, the extended symmetries necessarily have non-trivial center. The
maximal effective and transitive subgroup is therefore projective. When the
resulting central charges are introduced, the discrete part of the group
leads to the infinite Heisenberg algebra and the dilatational part leads to
the Virasoro algebra. These algebras govern tangent vectors when extended
groups are gauged, leading to nonvanishing commutators between canonically
conjugate fields.

Also, extended scaling groups fill some gaps in the previous formulation of
scale invariance. By formalizing the use of conformal weights as part of the
initial symmetry, the covariant derivative and gauge properties are
automatically appropriate to the weights of fields.

\bigskip

\noindent \textbf{Acknowledgement} \textit{The author thanks Andr\'{e} Wehner for his careful reading of this manuscript.}

\bigskip

\noindent \textbf{Appendix A: Scale-covariant weight maps}

\medskip
Here we find the maximal scale-covariant subgroup of the automorphism group of the set of integer-valued conformal weights.

Let $\Phi $ be the automorphism group of the integers, $\Phi =\{\varphi \ |\
\varphi :Z\rightarrow Z,bijective\}$. The set $L=\{(length)^{m}\ |\ m\in Z\}$
provides a representation $\Phi _{L}$ of $\Phi $ by defining $J_{\varphi
}:L\rightarrow L$ as 
\begin{equation}
J_{\varphi }l^{k}=l^{\varphi (k)}
\end{equation}
We seek the maximal scale-covariant subgroup $\Phi _{D}$ of $\Phi _{L},$ in
the sense that the action of dilations, $e^{\lambda D},$ should be
well-defined on $\Phi _{D}.$ That is, $\Phi _{D}$ must be closed under the
action of dilatation: 
\begin{equation}
e^{\lambda D}J_{\varphi }e^{-\lambda D}=\sum_{\varphi ^{\prime }}f_{\varphi
\varphi ^{\prime }}J_{\varphi ^{\prime }}\in \Phi _{D}  \label{Covariance}
\end{equation}
for all $J_{\varphi }$ in $\Phi _{D}.$ Consider the action of eq.(\ref
{Covariance}) on $L.$ We compute, for all $k,$ 
\begin{eqnarray}
e^{\lambda D}J_{\varphi }e^{-\lambda D}l^{k} &=&\sum_{\varphi ^{\prime
}}f_{\varphi \varphi ^{\prime }}J_{\varphi ^{\prime }}l^{k} \\
e^{\lambda \varphi (k)-\lambda k}l^{\varphi (k)} &=&\sum_{\varphi ^{\prime
}}f_{\varphi \varphi ^{\prime }}l^{\varphi ^{\prime }(k)}
\end{eqnarray}
which is satisfied iff both 
\begin{eqnarray}
f_{\varphi \varphi ^{\prime }} &=&e^{\alpha }\delta _{\varphi \varphi
^{\prime }} \\
\lambda \varphi (k)-\lambda k &=&\alpha (\varphi )
\end{eqnarray}
with $\alpha (\varphi )$ independent of $k.$ Thus, for all elements of the
subgroup, $\varphi =\varphi ^{\prime }$ and $\varphi (k)=k+\alpha (\varphi
)/\lambda \in Z.$ Therefore, the action of $\varphi $ is characterized by a
single integer $-m=$ $\alpha (\varphi )/\lambda \in Z$ and is given
explicityly by $\varphi (k)=k-m.$ Labelling $J_{\varphi }$ as $J_{m},$ we
have 
\begin{equation}
e^{\lambda D}J_{m}e^{-\lambda D}=e^{-\lambda m}J_{m}
\end{equation}
or infinitesimally, $\lbrack J_{m},D]=mJ_{m}$.

We conclude that the set $\Phi _{D}=$ $\{J_{m}|J_{m}l^{k}=l^{k-m},m\in Z\}$
with $J_{m}$ satisfying 
\begin{eqnarray}
\lbrack J_{m},J_{n}] &=&0 \\
\lbrack J_{m},D] &=&mJ_{m}
\end{eqnarray}
is the maximal covariant subset. The set is easily seen to form a subgroup
since $J_{0}$ is the identity and $J_{k}$ has inverse $J_{-k}.$ Also note
that the set $\{D,J_{m}\}$ is the basis for an infinite Lie algebra.

\bigskip

\noindent \textbf{Appendix B: Nonvanishing Jacobi relations }

\medskip The non-vanishing Jacobi relations for the tangent commutator
algebra involve only the structure constants $c_{\alpha \beta }^{\quad 0}$
and $c_{0\alpha }^{\quad 0}.$ For the homothetic algebras, both of these
vanish, while for the conformal group, $c_{0\alpha }^{\quad 0}=0.$ For a
general scaling algebra, the Jacobi identities are replaced by the following
Jacobi relations: 
\begin{eqnarray}
\lbrack G_{\alpha }^{k},[G_{\beta }^{m},G_{\gamma }^{n}]]_{et\ cyc}
&=&(nc_{\alpha \beta }^{\quad 0}\delta _{\gamma }^{\rho }+kc_{\beta \gamma
}^{\quad 0}\delta _{\alpha }^{\rho } \\
&&+mc_{\gamma \alpha }^{\quad 0}\delta _{\beta }^{\rho })\ G_{\rho }^{k+m+n}
\\
\lbrack G_{\alpha }^{k},[G_{\beta }^{m},L^{n}]]_{et\ cyc} &=&nc_{\alpha
\beta }^{\quad 0}L^{k+m+n} \\
&&+(mc_{0\alpha }^{\quad 0}\delta _{\beta }^{\rho }-kc_{0\beta }^{\quad
0}\delta _{\alpha }^{\rho })G_{\rho }^{k+m+n} \\
\lbrack G_{\alpha }^{k},[G_{\beta }^{m},J_{n}]]_{et\ cyc} &=&nc_{\alpha
\beta }^{\quad 0}J_{k+m+n} \\
\lbrack G_{\alpha }^{k},[L^{m},L^{n}]]_{et\ cyc} &=&(m-n)\ c_{0\alpha
}^{\quad 0}L^{k+m+n} \\
\lbrack G_{\alpha }^{k},[L^{m},J_{n}]]_{et\ cyc} &=&nc_{0\alpha }^{\quad
0}J_{k+m+n}
\end{eqnarray}
For the homothetic algebras, the usual Jacobi identities hold, so that the
extended homothetic algebra is an infinite dimensional Lie algebra. For the
conformal group (with a similar simplification for any Lie algebra with a
definite weight  basis, so that $c_{0\alpha }^{\quad 0}=0)$ the
non-vanishing Jacobi relations are 
\begin{eqnarray}
\lbrack P_{a}^{k},[K_{b}^{m},G_{\gamma }^{n}]]_{et\ cyc} &=&2(n\eta
_{ab}G_{\gamma }^{k+m+n} \\
&&+kc_{b\gamma }^{\quad 0}P_{a}^{k+m+n}+mc_{\gamma a}^{\quad
0}K_{b}^{k+m+n})\  \\
\lbrack P_{a}^{k},[K_{b}^{m},L^{n}]]_{et\ cyc} &=&2n\eta _{ab}L^{k+m+n} \\
\lbrack P_{a}^{k},[K_{b}^{m},J_{n}]]_{et\ cyc} &=&2n\eta _{ab}J_{k+m+n}
\end{eqnarray}

\end{document}